\documentstyle[preprint,prb,aps]{revtex}
\begin{document}
\draft
\title{Interesting magnetic properties of Fe$_{1-x}$Co$_x$Si alloys}
\author{M. K. Chattopadhyay, S. B. Roy and Sujeet Chaudhary}
\address{Low Temperature Physics Laboratory,
Centre for Advanced Technology,Indore 452013, India}
\date{\today}
\maketitle
\begin{abstract}
Solid solution between nonmagnetic narrow gap semiconductor FeSi
and diamagnetic semi-metal CoSi  gives rise to interesting metallic
alloys with long-range helical magnetic ordering,
for a wide range of intermediate concentration. 
We report  various interesting magnetic properties of these alloys,
including low temperature re-entrant spin-glass like behaviour 
and a novel inverted magnetic hysteresis loop. 
Role of Dzyaloshinski-Moriya interaction  in the magnetic response 
of these non-centrosymmetric alloys is discussed.
\end{abstract}                          

The narrow-gap semiconductor FeSi has drawn attention of
condensed matter physicists repeatedly 
since late nineteen thirties \cite{1}. The revival 
of strong interest \cite{2} in FeSi during last  decade is 
mainly due to its similarities with those of narrow gap rare-earth 
intermetallics popularly known as "Kondo insulators " 
\cite{3}.  This comparison gives rise to the possibility of the study of
complex many-body phenomena associated with Kondo-lattice systems.
Doping with Al in FeSi leads to a heavy fermion 
metal through a metal-insulator transition with strong 
similarities with that for Si:P (Ref.4) with the exception
of a strongly renormalized effective carrier mass.
The Fe$_{1-x}$Co$_x$Si alloys are also remarkable in that 
they are magnetic for almost all the intermediate concentration 
regime \cite{5,6,7}, while the end compounds FeSi and CoSi 
are nonmagnetic, the  latter being  a diamagnetic 
semi-metal\cite{8}. The recent discovery of unusual positive 
magnetoresistance\cite{7} in these supposedly helimagnetic 
Fe$_{1-x}$Co$_x$Si alloys\cite{6} alongwith the suggestion
of the interplay of quantum   
coherence effect at relatively high temperature are quite exciting. 
The unusual square-root field-temperature dependence of electrical 
conductivity and the positive nature of the magnetoresistance are correlated to 
square-root singularity in the density of states probably 
associated with "enhanced electron-electron interactions" in 
a disordered ferromagnet with low carrier                   
concentration\cite{7}. These results suggest a possible new microscopic
mechanism of magnetoresistance that could lead to the 
development of new type of magnetic materials of technological
importance\cite{9}.  In the light of 
these unusual findings, we became motivated for a closer 
scrutiny of the magnetic properties of  Fe$_{1-x}$Co$_x$Si 
alloys, especially in the low field and low temperature 
regime. There exist already some hints of unusual 
low field magnetic properties of Fe$_{1-x}$Co$_x$Si alloys in
the form of an almost singular behaviour in magnetization and 
cusp-like minimum in magnetoresistance near H = 0 (Ref. 7). 
In this communication we report results of high resolution 
magnetization measurements in Fe$_{1-x}$Co$_x$Si alloys
highlighting (i) low temperature  low field  
re-entrant spin-glass like behaviour (ii) interesting 
thermomagnetic history effects including a novel "inverted 
hysteresis loop" with negative remanence.
The observation of this latter effect (which  was so far
considered to be limited to thin-film type 
of magnetic materials \cite{10,11}) in  relatively simple alloys like 
the present (Fe,Co)Si,  is definitely interesting. 
We shall argue that the occurrence of Dzyaloshinski-Moriya interaction in 
the  present non-centrosymmetric cubic B20 
Fe$_{1-x}$Co$_x$Si alloys\cite{6} is playing an important role for the 
observed magnetic properties.

The polycrystalline samples of  Fe$_{1-x}$Co$_x$Si; x = 0.1, 
0.15, 0.35, 0.45 and 0.65 were prepared by argon arc         
melting from high purity starting materials. The samples 
were annealed for 90 hours in vacuum at 900$^0$C for improving  
the homogeneity. Magnetiztion measurements were performed using 
a commercial SQUID-magnetometer (Quantum Design, MPMS-5). A 
scan-length of 4 cm with 32 data points in each scan was 
used for the measurements. However, all the important 
results were checked by varying the scan-length from 2 to 8
cm, to rule out any possible role of the small field 
inhomogeneity of the superconducting magnet (which is 
actually scan-length dependent) in the observed magnetic    
properties. Also before the start of each experimental 
cycle the sample chamber is heated to 200 K and flushed 
with helium; this is to get rid of any oxygen leaking into  
the sample chamber over a period of time.

In Fig.1(a) we plot magnetization (M) and inverse dc-susceptibility
($\chi ^{-1}$) versus temperature (T)   
for  Fe$_{1-x}$Co$_x$Si with x = 0.15 and 0.35.
Estimated Curie temperatures
(T$_C$) agree well with those reported in the literature \cite{7}.
In Fig.1(b)-(c) we plot M vs field (H) plots for these alloys at various
T both below and above T$_C$. 
Data also exist for x = 0.1 and 0.45 but not shown here for 
the sake of clarity and conciseness. The almost singular behaviour
in M(H) near H = 0 for T $<$ T$_C$ as reported in Ref. 7 is quite evident 
in Fig. 1(b)-(c). We shall now concentrate in the low H magnetic response 
of these alloys. In fig.2 we present M vs T plots for x = 0.35 alloy obtained
both in the zero field cooled (ZFC) and field cooled (FC) mode in various
applied H. We observe two distinct features  for H $\leq$ 500 Oe, namely  
(1) a peak in M$_{ZFC}$(T) and a sharp change in slope in M$_{FC}$(T)
at a temperature T$_P$ ($<$ T$_C$).
(2) a distinct thermomagnetic irreversibility (TMI) i.e. 
M$_{ZFC} \neq$ M$_{FC}$ for T$\leq$ T$_P$.
Same qualitative features have also been observed for x = 0.1,0.15 and 0.45.
Both these features, which disappear with H $>$ 500Oe,
have not been reported so far (to our knowledge) for these (Fe, Co)Si alloys.

The low-T low-H magnetic response described above has appreciable resemblance with 
the re-entrant spin-glasses \cite{12,13}. To investigate more in this regard
we have studied the H dependence of magnetization in details
in two different T-regimes : (1) T $<$ T$_P$, and (2) 
T$_P < $ T $<$ T$_C$. In Fig.3 we plot M vs H  for x = 0.35 alloy at 4.5K  
highlighting the following striking features: 
\begin{enumerate}
\item There is a distinct bulge in the 
virgin M-H curve obtained after zero-field cooling the 
sample from T $>$ T$_C$. This feature takes the virgin M-H curve  
in a limited H-regime outside the field descending (ascending) M-H 
curve obtained after field cycling to 50 kOe (-50 kOe).
\item In the field cycling process if the maximum field of    
excursion H$_{max}$ goes beyond the technical saturation 
point H$_{sat} (\approx$ 1 kOe at T = 4.5K), the M-H curve takes the shape 
of an inverted hysteresis loop, i.e., the descending field 
leg of the M-H curve lies below that of
the ascending field leg with positive coercivity and       
negative remanence (see the lower inset of Fig.3)\cite{14}. 
\item If  H$_{max}$ is limited to  H $<<$ H $_{sat}$,
M remains perfectly reversible. However, as
H$_{max}$ enters the H-regime where the virgin M-H curve
starts showing the non-linear behaviour in the form of a bulge,
a small but distinct positive hysteresis is observed (see
the upper inset of Fig.3). This hysteresis disappears 
as H approaches H = 0 in the descending field cycle and M
merges with the virgin M-H curve.
With  H$_{max}>$ H$_{sat}$ 
this positive hysteresis changes sign giving rise to 
an "inverted hysteresis loop" in the low field regime (H $<$ H$_{sat}$) while
the M-H curve remains perfectly reversible (within our experimental
resolution) in the high field regime (H $>$ H$_{sat}$).
\end{enumerate}

In the T-regime T$_P <$ T $<$ T$_C$ the bulge in the virgin M-H curve 
and the associated positive hysteresis are not observed. However, inverted
hysteresis loop behaviour continues to exist at H $<$ H$_{sat}$) 
even for T $>$ T$_P$. And as before, the M-H curves remain reversible for
H $>$ H$_{sat}$. 
All these features of the M-H curve are also observed 
in x = 0.1, 0.15 and 0.45 alloys in the same qualitative manner.

The observed peak in M$_{ZFC}$(T) and TMI in M-T plots in 
Fig.2 with H $\leq$ 500 Oe 
can naively be interpreted in terms of the hindrance of 
domains' motion in a ferromagnetic system\cite{15}. 
However, even if the various anomalous aspects of the M-H
curves decsribed above are ignored, the estimated coercivity 
field $|$H$_C|$  of the order of 15 Oe in our x = 0.35 alloy at 
T = 4.5K rules out such a simple explanation in our measurements
with applied H of 500 Oe which is much larger than  $|$H$_C|$ . 
Moreover the distinct change in slope 
in M$_{FC}$(T) cannot be associated with any domain-related phenomena.  
These results suggest that there exist probably a 
re-entarnt spin-glass like  magnetic phase\cite{13} for T $<$ T$_P$ in these 
alloys. This low T phase appears to be quite fragile 
and can easily be erased with moderate applied      
magnetic field. It is interesting to note here that the 
anomalous bulge in the virgin M-H curve 
is observed below T$_P$ only, and it is
quite clear from the above arguments that it is not associated with any domain related 
phenomenon either. We suggest that this non-linear beahviour in the
virgin M-H curve 
probably represents a field-induced transition from a 
low-H magnetic state to a high-H one. 
The bulge in the virgin M-H curve has been reported earlier for (Fe,Co)Si 
in passing \cite{5,7}, and in the absence of a detailed     
magetization study it was attributed to domain related 
effects in a ferromagnet \cite{5}. A similar anomalous behaviour of the 
virgin M-H curve in CeFe$_2$-based pseudobinary alloys has been associated 
recently with the first order nature of a field induced metamagnetic 
transition \cite{16}.

The question now arises how to rationalise the interesting 
magnetic properties of (Fe,Co)Si within the framework 
already developed for these alloys. Small angle neutron 
scattering measurements \cite{6} have suggested the          
magnetic ordering in (Fe,Co)Si alloys to be of long period  
helimagnetic in nature. A model to explain such long 
period helimagnetic order can be based on a competition     
between a Dzyaloshinski-Moriya (DM) interaction and a 
Heisenberg type exchange interaction \cite{5,6}. The          
non-centrosymmetric cubic B20 structure of (Fe,Co)Si alloys 
supports the existence of DM interaction. Can this 
competition between  these two types of interactions in
(Fe,Co)Si alloys  give rise to a re-entrant 
spin-glass like behaviour ? DM interaction apparently plays 
an important role in metallic spin-glasses and re-entrant  
spin-glasses \cite{13}. In this context the occurrence 
of a re-entrant spin-glass like phase in (Fe,Co)Si 
alloys is not entirely unexpected, especially with the
presence of inherent disorder in the (Fe,Co) sublattice. In 
fact hints of repartition of the magnetic moments in the 
helix due to alloying effects exist in early neutron 
studies \cite{6}. Satellites due to both clockwise and 
counterclockwise helixes were observed in neutron 
measurements in zero field cooled samples . After 
excursion to a high H, the single clockwise helix was 
stabilized to the field direction with no satellites 
observed in any other direction \cite{6}. On reduction of H to zero 
the helix does not come back to a specific eqilibrium 
direction. This is in contrast to the case of isostructural 
ordered compound MnSi where also the helix follows the field 
but comes back to $<111>$ direction in low H (Ref.6). It was 
argued that the disorder in (Fe,Co) sublattice caused a 
local fluctuations of the co-efficient of D-M interaction to
produce two kinds of domains consisting of either a 
clockwise or counterclockwise helix in the zero field cooled 
state. The local fluctuation of magnetization might play a 
role of pinning effect of the    
magnetic impurity preventing the propagation vector from pointing to the 
equilibrium direction \cite{6}.

The observed "inverted hysteresis loop", however, does not 
find a simple explanation within the above framework.
Such "inverted hysteresis loops" have been observed in recent
years in specific exchange-coupled multilayers such as 
Co/Pt/Gd/Pt and epitaxial Fe films on W(001) (Ref. 10 and 11). In such
materials their thin film structure apparently play an important role
and hence it is considered that "inverted hysteresis loop" is 
probably a phenomenon limited to thin-film type of magnetic materials.
However, there is a very recent report of "inverted hysteresis loop" in 
a bulk magnetic material comprising of cyanide-bridged multi metal
complexes \cite{17}. The observed "inverted hysteresis loop" in this 
bulk material is explained "by the competition between the sublattice 
magnetization rotation  due to the spin-flip transition and the 
trapping effect due to the uniaxial magnetic anisotropy" \cite{17}.
While there exists signature as discussed above of spin-flip transition in the present 
(Fe,Co)Si alloys and also the suggestion that D-M interaction can cause
trapping effect for domains  especially if spins are canted
within the domains \cite{13}, it is a bit premature 
to import the similar picture here. More experimental information, 
especially the microscopic ones like neutron scattering, is required 
to form even a qualitative model to expain the "inverted hysteresis loop"
in the present system.

We note in Figs.1 (b)-(c) that while the technical saturation point is reached in the   
M-H curves below T$_C$ for x = 0.15 and 0.35 alloys 
at fairly low fields (H$_{sat} \approx$1 kOe),
M actually continues to increase beyond H$_{sat}$ even 
up to the highest field of our measurement i.e. 50 kOe. This two
stage magnetization process indicates that after the initial 
low-H alignment, the local 
spins, which are probably canted, line up slowly with  
further increase in H beyond H$_{sat}$. We can actually make 
a reasonable fit of the M-H curve in the regime H$_{sat} <$ H $<$ 50 kOe to a
H$^{1/2}$ behaviour.  Similar behaviour has also been 
observed for x = 0.1 and 0.45 alloys. Manyala et al \cite{7} have earlier reported 
that magnetoresistance in some of these alloys also varied as H$^{1/2}$ in the H regime
beyond technical saturation. This clearly indicates that the behaviour of 
these alloys is quite different from a conventional ferromagnet even 
in the high-H regime .

In conclusion, our present dc-magnetization measurements in conjunction with
the results of earlier neutron studies \cite{6}, suggest that there 
exists a low-T low-H magnetic state in (Fe, Co)Si alloys 
which resembles a lot of the re-entrant spin glasses. With the incease in T and H, it transforms to a presently recognized high-H high-T helical FM
state. Careful neutron measurements in various (H,T) regimes with 
different thermomagnetic history will be useful to settle this issue.
The high-T high-H magnetic state of these alloys has an
unusual magnetic field dependence in the form of M $\propto$ H$^{1/2}$.
Also, the magnetization response is reversible above the field for 
technical saturation H$_{sat}$, and produces a narrow "inverted hysteresis loop" 
below H$_{sat}$. A proper understanding of these magnetic responses and their 
possible correlation to the technologically promising magnetotransport\cite{7,9}
will help in the search for newer magnetic materials tunable for practical
use.  
  
\section{Acknowledgement}
We would like to acknowledge Dr. K. J. Singh for the help in sample 
preparation and Dr. P. Chaddah for useful discussion.

\begin{figure}
\caption{(a) M and ($\chi^{-1}$) vs T plots,  (b) and (c) M vs H plots 
for (Fe$_{1-x}$Co$_x$)Si, x = 0.15 and 0.35. In Fig. 1 (a) M is obtained with a
field of 2 kOe and $\chi^{-1}$ from magnetization obtained with H =  200 Oe}
\end{figure}
\begin{figure}
\caption{M vs T plots for (Fe$_{0.65}$Co$_{0.35}$)Si obtained both
in the ZFC and FC mode with H = 200 Oe, 500 Oe, 2 kOe and 20 kOe.}
\end{figure}
\begin{figure}
\caption{M vs H for (Fe$_{0.65}$Co$_{0.35}$)Si at T = 4.5 K highlighting various
anomalous features of the M-H curve. (i) Below H$_{sat}$ the M-H loop is
inverted in nature i.e the ascending field M-H curve (diamond) is lying
above the descending field  M-H curve (square) . This gives rise to a negative remanance which is highlighted 
in the lower inset. Above H$_{sat}$ the M-H curve is reversible. (ii) In certain H regime the virgin M-H curve (circle ) is lying outside
the enevelope curves. Minor hysteresis loops (MHL) drawn from the non-linear regime
of the virgin curve (but the maximum field of excursion H$_{max}$ being lower than H$_{sat}$) show positive hysteresis but merge with the virgin
curve again before reaching H = 0. MHL's drawn from the low field linear regime of 
the virgin curve are perfectly reversible ( see the upper inset).}
\end{figure}   

\begin{references}
\bibitem{14}To rule out that the observed "inverted hysteresis loop" is not an
experimental artefact associated with our magnetometer, we have measured the
M-H loop of a ferromagnetic CeFe$_2$ (T$_C \approx$235 K)sample of similar signal strength at 4.5 K.
It shows a narrow but usual hysteresis loop with a positive remanence and
a negative coercivity field  of about 15 Oe.
\end{references}
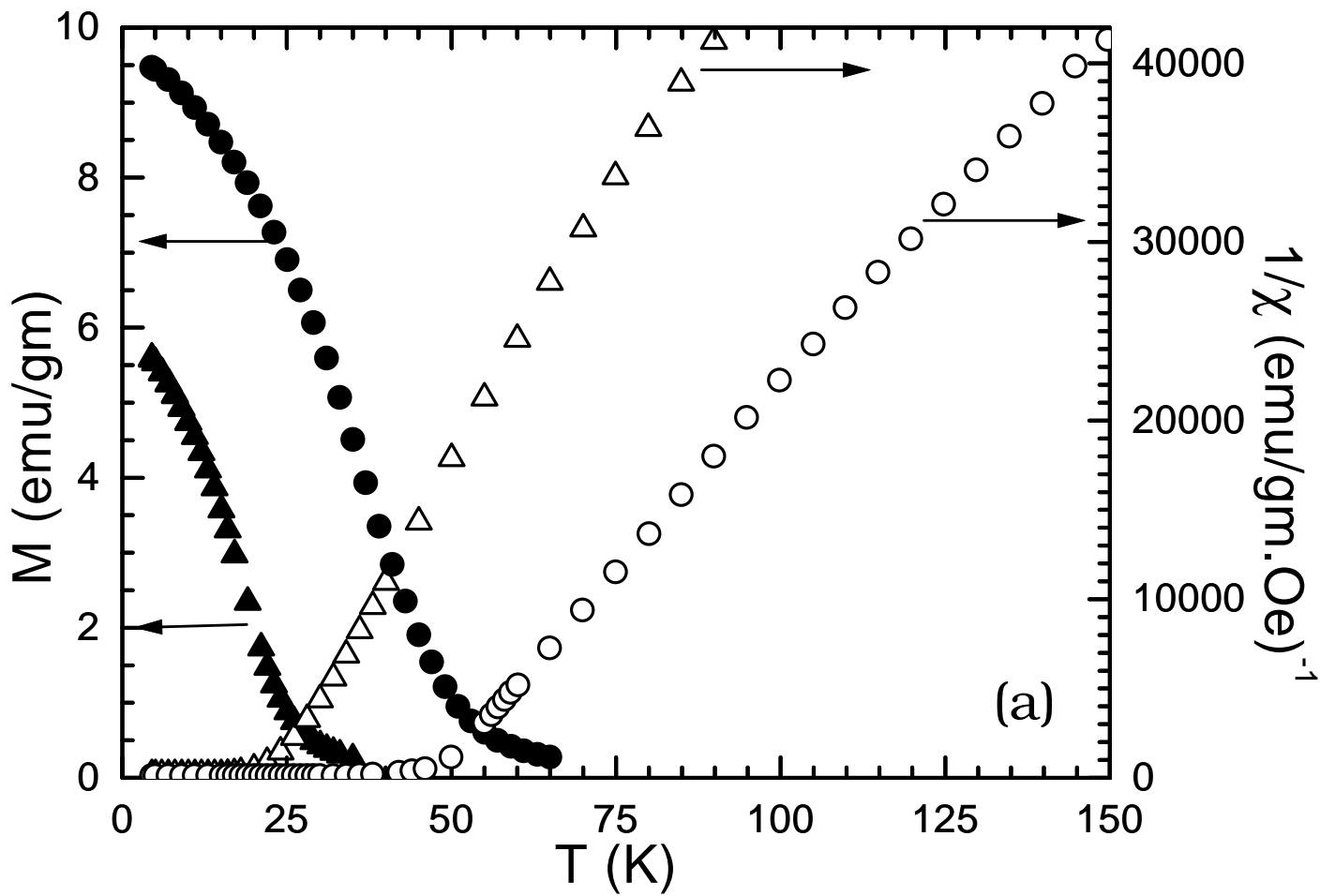
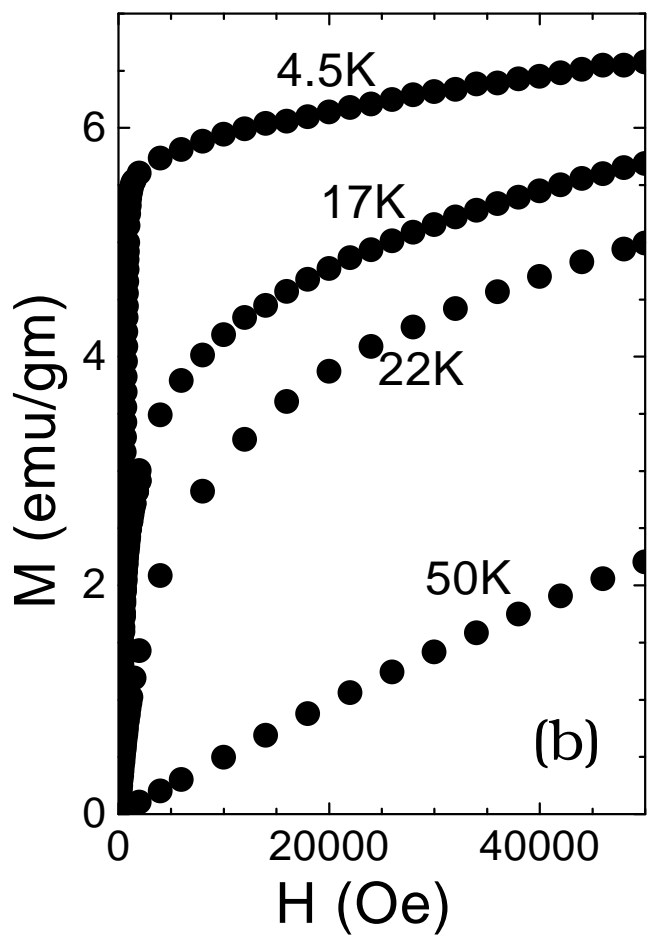
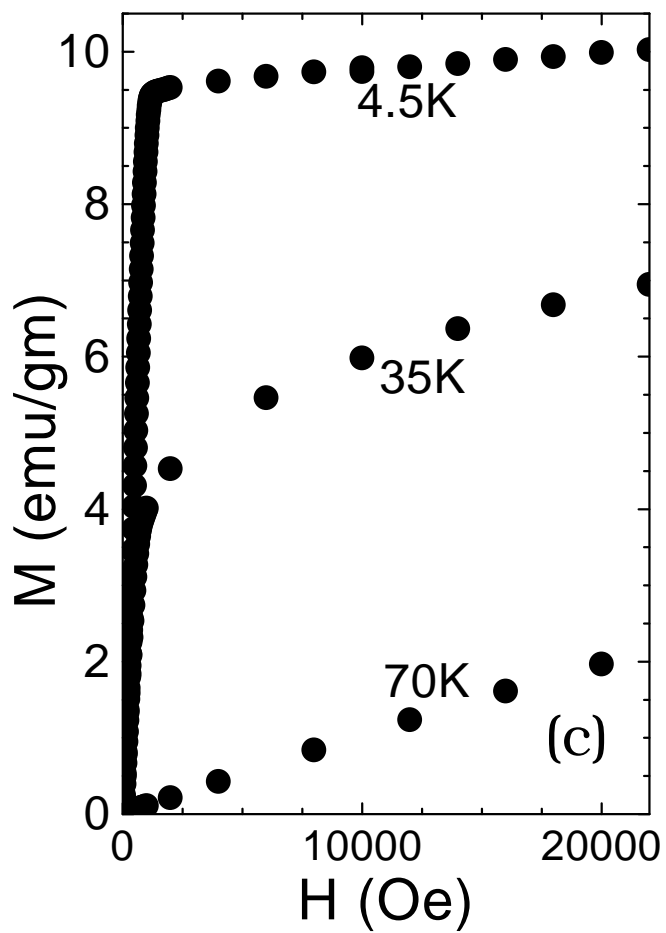
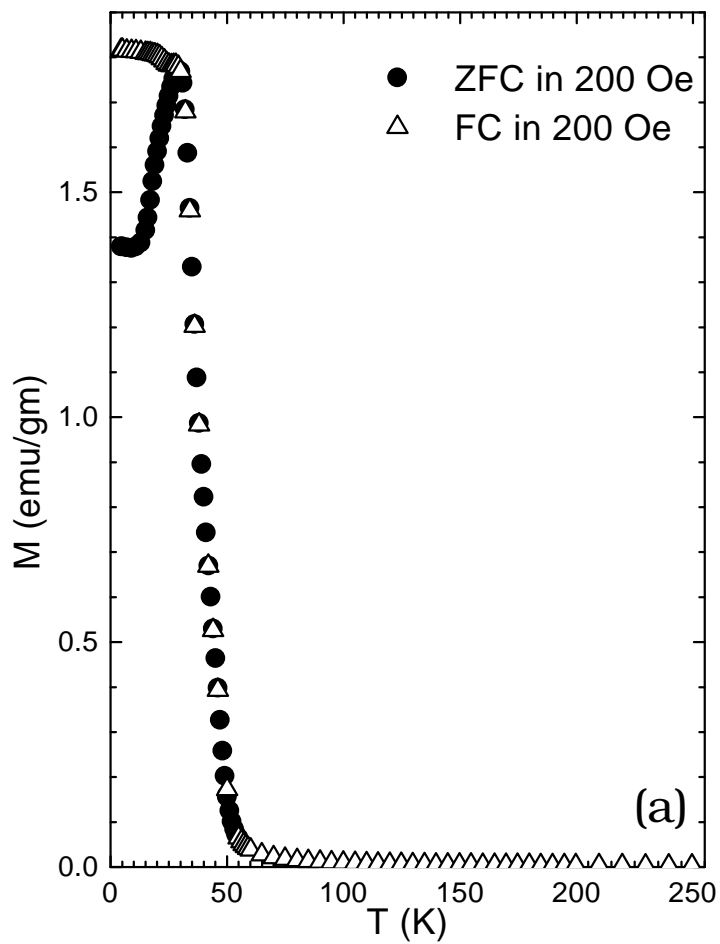
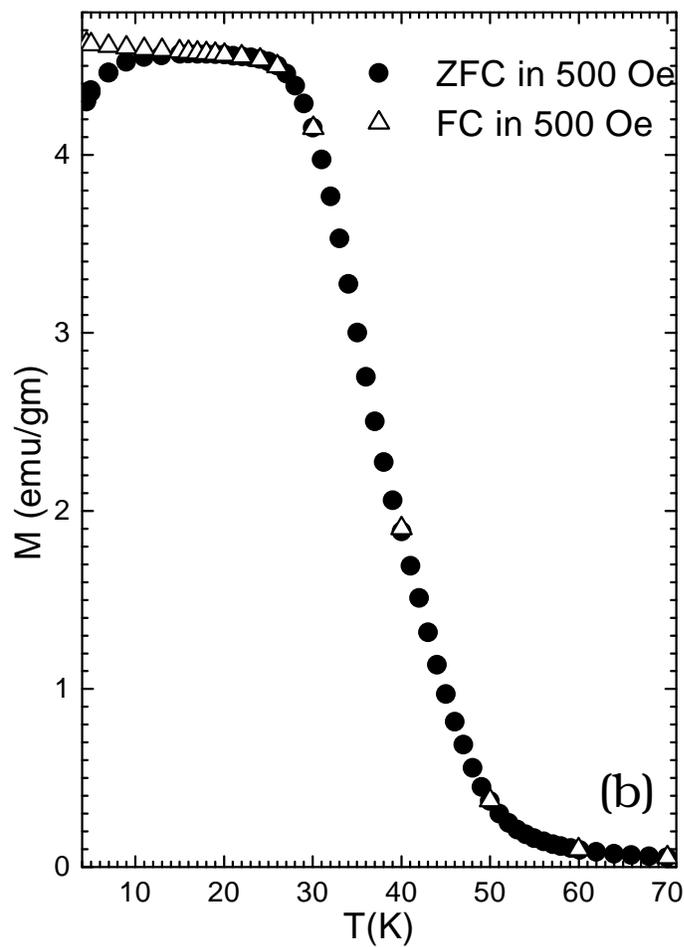
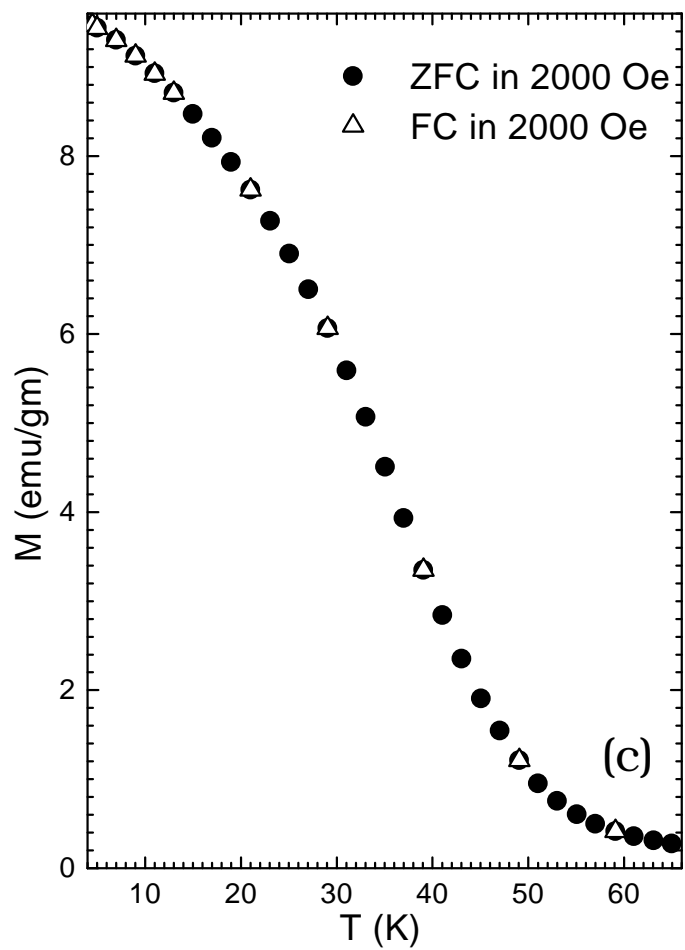
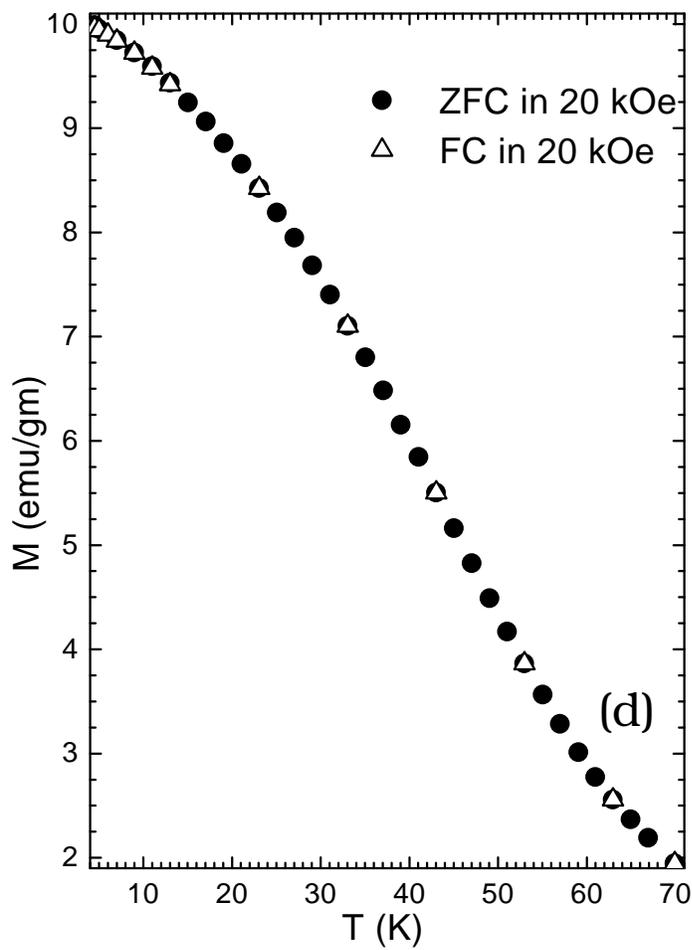
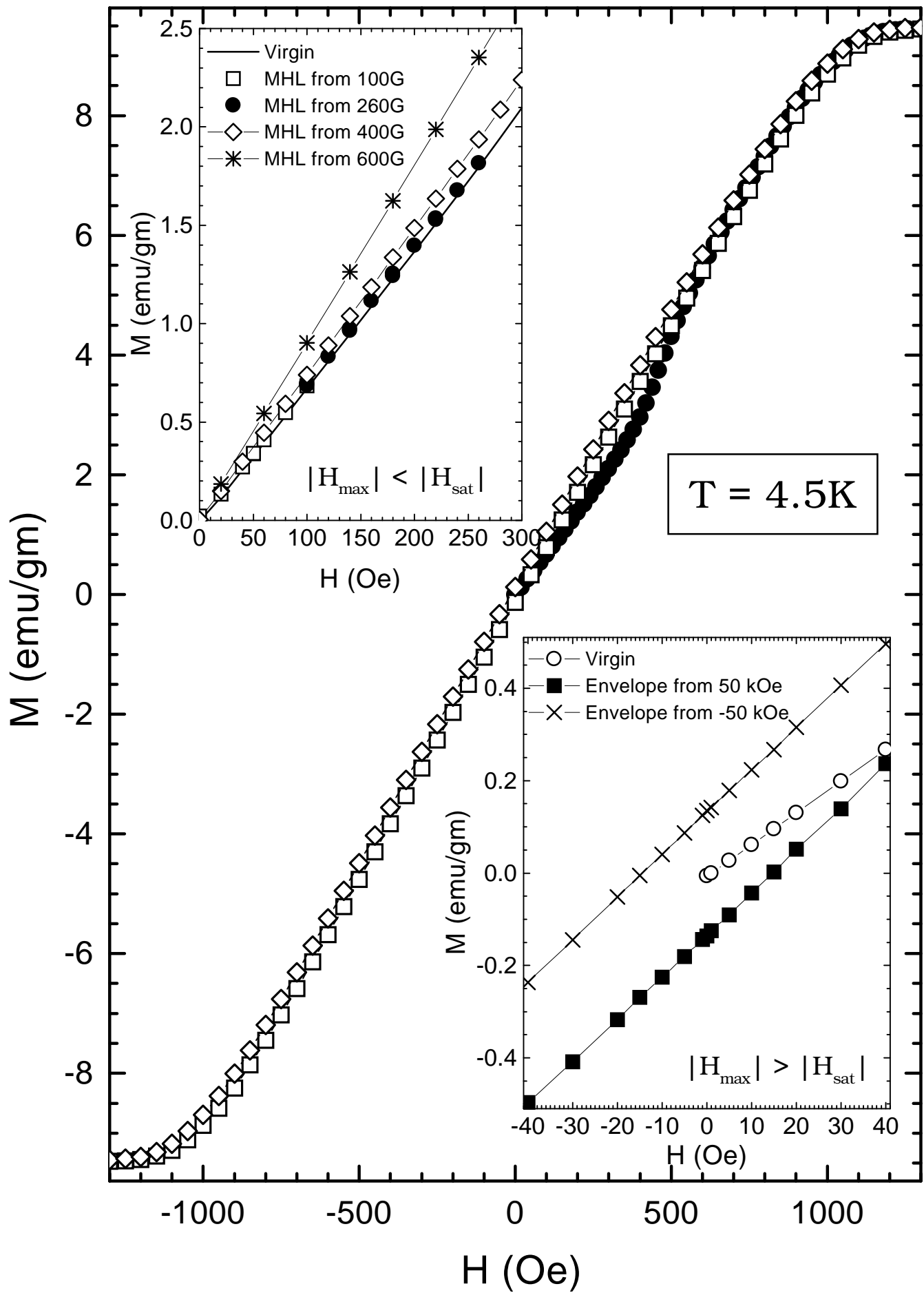
\end{document}